\documentclass[aps,prl,showpacs,twocolumn]{revtex4}
\usepackage{graphicx}
\usepackage{dcolumn}
\usepackage{xspace}

\begin{document}

\title{Two-speed phase dynamics in Si(111) (7$\times$7)-(1$\times$1) phase transition}

\author{Ye-Chuan Xu}
\affiliation{Institute of Physics, Chinese Academy of Sciences,
Beijing 100080, China} \affiliation{Beijing National Laboratory
for Condensed Matter Physics, Beijing 100080, China}
\author{Bang-Gui Liu}\email{bgliu@mail.iphy.ac.cn}
\affiliation{Institute of Physics, Chinese Academy of Sciences,
Beijing 100080, China} \affiliation{Beijing National Laboratory
for Condensed Matter Physics, Beijing 100080, China}

\date{\today}

\begin{abstract}
We propose a natural two-speed model for the phase dynamics of
Si(111) 7$\times$7 phase transition to high temperature
unreconstructed phase. We formulate the phase dynamics by using
phase-field method and adaptive mesh refinement. Our simulated
results show that a 7$\times$7 island decays with its shape kept
unchanged, and its area decay rate is shown to be a constant
increasing with its initial area. LEEM experiments concerned are
explained, which confirms that the dimer chains and corner holes
are broken first in the transition, and then the stacking fault is
remedied slowly. This phase-field method is a reliable approach to
phase dynamics of surface phase transitions.
\end{abstract}

\pacs{68.35.-p, 05.10.-a, 68.37.-d, and 05.70.-a}

\maketitle

{\it Introduction} It is well known that bulk terminated
1$\times$1 semiconductor surfaces usually are unstable against
reconstruction at low temperatures and some of the reconstructed
surface phases transit to the 1$\times$1 ones at high
temperatures\cite{01}. Si surfaces are well studied because
silicon plays the centering role in modern computer
industry\cite{02,03,04} and potential silicon-based
spintronics\cite{05}. The most important Si surface phase is the
Si(111) 7$\times$7 reconstructed surface \cite{das}. It is the
ground state structure of Si(111) surface and transits to an
unreconstructed 1$\times$1 structure above $T_c$=1125
K\cite{rr2,rr3,rr5,f0,f1,f3,f4,f5,f6,s1}. {\it In situ}
scanning-tunneling-microscopy (STM) observations\cite{rr2,rr3},
supported by Monte Carlo simulations\cite{tMC1,f4}, indicated that
the stacking faulted triangle unit is the basic building block in
forming the 7$\times$7 surface phase. The phase transition is
believed to be of first order\cite{l1,f1,f3,f4,f5,f6,l3,tMC1},
although earlier data implied a continuous phase
transition\cite{s1}. It is of much interest to clarify the phase
dynamics during the phase transition. Low-energy electron
microscopy (LEEM) imaging is a powerful approach to experimentally
measure time-dependent surface phases\cite{l2,l3,l4}. Recent LEEM
results showed that big 7$\times$7 islands always decay faster
than small ones, with area decay rates increasing with their
initial areas\cite{l3}. It seams like that the area decays have
something like momentum. This phenomenon is very intriguing. It
means that the phase transition dynamics and the microscopic
processes meanwhile are very much complex, what we have known
about them may be just a tip of the iceberg, and therefore much
more investigation is in need.

Here we investigate the time-dependent phase dynamics during the
phase transition of Si(111) 7$\times$7 islands to the
unreconstructed high temperature phase. Through analyzing
experimental results and the atomic structures of both surface
phases, we propose a two-speed model for the phase dynamics. We
formulate the phase dynamics using phase field method, famous for
various growth and solidification
issues\cite{pf1,pf2,pf2a,pf3,pf4,g1,g2}, and adaptive mesh
refinement technique\cite{pfa}. Our simulated results indicate
that the dimer chains and corner holes are broken first in the
phase transition, and then follows the slower remedying of the
stacking fault. Our simulated images and linear area decays of
7$\times$7 islands are in agreement with the experimental results,
which supports the two-speed model. This is a reliable approach to
understand the phase dynamics of the Si(111) 7$\times$7 phase
transition and other surface phase transitions.

{\it Experimental clues} The bulk terminated Si(111) 1$\times$1
surface and the dimers-adatoms-stacking-fault (DAS) model of
Si(111) 7$\times$7 reconstructed surface phase are shown in Fig. 1
\cite{das,01}. This 1$\times$1 surface is not experimentally
realized. The real experimental phase at high-temperature is a
``1$\times$1'' phase that is formed by covering the bulk
terminated 1$\times$1 surface with 0.25 monolayer (ML) of
fast-moving adatoms\cite{r4a,r4b}. The
7$\times$7$\rightarrow$``1$\times$1'' phase transition needs extra
0.17 ML of adatoms because the former has 0.08 (4/49) ML more
adatoms than the bulk terminated 1$\times$1 surface. On the
experimental side, the appearing of the 7$\times$7 LEED pattern,
or the seventh-order spots, means that the
``1$\times$1''$\rightarrow$7$\times$7 transition happens, and on
the other hand the absence of it means that the 7$\times$7
structure is destroyed\cite{r22}. The adatoms stay at T$_4$ sites
above the second-layer atoms in the 7$\times$7 surface, and some
of them moves to H$_3$ sites above the forth-layer atoms when the
surface transits to the ``1$\times$1'' phase\cite{r21}. The
movement of adatoms from T$_4$ sites to H$_3$ sites or vice versa
is easy because it can be realized without breaking any bond. The
formation of the dimer chains, corner holes, and stacking fault is
considered essential to the 7$\times$7 reconstruction\cite{r13}.
The presence of the LEEM image means that the key 7$\times$7
factors, the dimer chains and corner holes, still exist. Without
the key factors, the 7$\times$7 phase is destroyed. The resultant
region without dimer chains and corner holes, although still
having stacking-fault, cannot be distinguished from the
``1$\times$1'' phase by means of LEEM experiment, being `ghost'
region to LEEM imaging\cite{r24}.
\begin{figure}[tb]
\includegraphics[width=8.5cm]{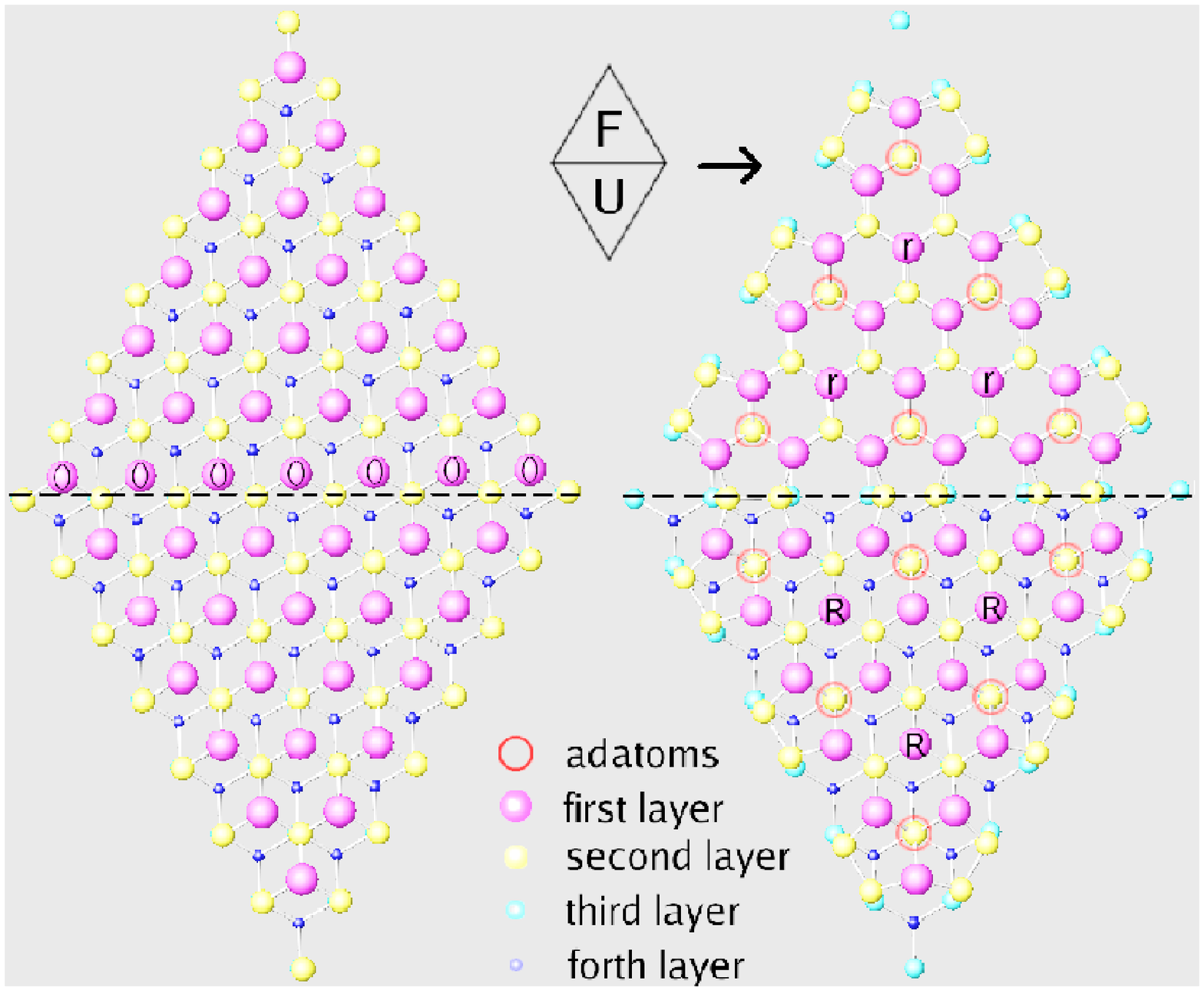} \caption{(color online).
The atomic structures of bulk terminated Si(111) 1$\times$1
surface (left) and the DAS model of Si(111) 7$\times$7
reconstructed surface (right). The top four layers are shown for
both structures, and in addition, the 12 adatoms are included for
the latter. The 7$\times$7 unit is divided into the faulted (F,
upper) half and the unfaulted (U, lower) half by the dash line.
}\label{fig1}
\end{figure}

{\it Model and phase-field realization} Therefore, we believe that
the atoms reorganization during the
7$\times$7$\rightarrow$``1$\times$1'' transition can be described
by two basic processes: (a) the destroying of the dimer chains and
corner holes, and (b) the remedying of the stacking-fault. The
latter process should take more time because it requires
collective movement of many atoms concerned. We adopt phase field
method\cite{pf1,pf2,pf2a,pf3,pf4} to simulate the phase dynamics
during the 7$\times$7$\rightarrow$``1$\times$1'' transition. We
use phase-field variables $\phi$ and $\xi$ together to describe
the 7$\times$7 phase with respect to the ``1$\times$1'' phase.
$\phi$ is used for the aspect of the dimers and corner holes
according to process (a) and $\xi$ for the aspect of the stacking
fault according to process (b). Both $\phi$ and $\xi$ are
functions of the time $t$ and two-dimensional space coordinates
($x$,$y$), and are made to have two stable values, 1 and -1, as is
done in usual phase-field simulations\cite{pf1,pf2,pf2a,pf3,pf4}.
The adatoms are described with another variable $u$ to be defined
in the following. Therefore, the complete 7$\times$7 region, with
not only the dimer chains and corner holes but also the
stacking-fault, is described by $\phi$=1 and $\xi$=1; and the
complete ``1$\times$1'' region by $\phi$=-1 and $\xi$=-1.
$\phi$=1, regardless of $\xi$, reflects the presence of the dimer
chains and corner holes, the key 7$\times$7 character detected in
LEEM imaging experiment. $\xi$=1 and $\phi$=-1 together describe
the temporary `ghost' regions during the 7$\times$7
--``1$\times$1'' phase transition. The governing equations can be
expressed as
\begin{equation}\left\{\begin{array}{l}
\displaystyle \tau_\phi \frac{\partial \phi}{\partial t}  =
w_\phi^2\nabla^2\phi - f_\phi(\phi) -\lambda g_\phi(\phi)u -
\lambda^\prime
g_{\phi}(\phi) g(\xi) \\[2mm]
\displaystyle \tau_\xi \frac{\partial \xi}{\partial t} =
w_\xi^2\nabla^2\xi - f_{\xi}(\xi) - \lambda^\prime g(\phi)
g_{\xi}(\xi)
\end{array}\right.\label{eq1}\end{equation}
where $\nabla^2$ is defined as $\frac{\partial^2 }{\partial
x^2}+\frac{\partial^2 }{\partial y^2}$, and the functions
$f_\eta(\eta)$ and $g_\eta(\eta)$ ($\eta$=$\phi$, $\xi$) are the
derivatives of functions $f(\eta)$ and $g(\eta)$, respectively. We
define $f(\eta)$ as $\exp[(\eta^2-1)^2]$, rather than usual
$(\eta^2-1)^2$, because the $\lambda^\prime$ terms are fifth-order
polynomials of $\phi$ and $\xi$. This choice not only keeps the
two stable values, +1 and -1, and the parabolic shape in their
neighborhoods without adding extra extreme values, but also
guarantees the stability of the system against the
$\lambda^\prime$ terms. The $\tau_{\phi}$ and $\tau_{\xi}$
describe the characteristic evolution times of the phase fields.
The $\nabla^2$ term and the time derivative one together
determines the evolution rate of the phase-field, with
$D_\eta$=$w_\eta^2/\tau_\eta$ acting as controlling parameter. In
addition, $w_\eta^2$ determines the transition zone of the
phase-fields in the asymptotic regime. Here we have $D_{\xi} <
D_{\phi}$ because process (b) is slower than process (a). The
$\lambda$ term describes the interaction between $u$ and $\phi$,
reflecting the growth or the shrinkage of LEEM-detectable islands
with the help of adatoms. The $\lambda^\prime$ term describes
enhanced evolution of $\phi$=1 ($\xi$=1) islands in the presence
of $\xi$=1 ($\phi$=1). We use $g(\eta)$=$(2+3\eta-\eta^3)/3$
because the growth and shrinkage of the $\phi$=1 region takes
place only by the $u$ exchange along its boundary and is enhanced
only with the existence of $\xi$=1 regions.

{\it Computational methods and parameters} The variable $u$
describes the local density difference of adatoms between the
``1$\times$1'' structure and the 7$\times$7 one. Because the
adatoms move very quickly, we suppose that $u$ does not directly
depend on space positions, but is a function of the phase field
$\phi$, $u$=$u_0 [1-\tanh(\phi)/\tanh(1)]/2$. This means that $u$
is zero deeply in the $\phi$=1 islands\cite{l3}, equal to $u_0$
deeply in the $\phi$=-1 regions, and in between in transition
zones where $\phi$ takes values between 1 and -1. We take
$u_0$=1.33nm$^{-2}$ in terms of experimental adatom
measurements\cite{r4b,01}. Other parameters are taken in terms of
basic time interval $dt$=0.08s and basic grid length $dx$=1.77 nm.
The latter is determined in terms of the area of the minimal
7$\times$7 equilateral triangle. Because the half 7$\times$7 unit
cell is an equilateral triangle, we conserve the threefold
anisotropy by using $w_{\phi}$=$w_1h(\theta_{\phi})$,
$w_{\xi}$=$w_2h(\theta_{\xi})$,
$\tau_{\phi}$=$\tau_1h^2(\theta_{\phi})$, and
$\tau_{\xi}$=$\tau_2h^2(\theta_{\xi})$. The function
$h(\theta_\eta)$ ($\eta$=$\phi$, $\xi$) is defined as
$h(\theta_\eta)$=$1+\epsilon \cos(3\theta_\eta)$, where $\epsilon$
parameters the anisotropy, and the directional function
$\theta_{\eta}$ is defined as
$\arcsin(-\eta_y/\sqrt{\eta^2_x+\eta^2_y})-\pi/2$, where the
$\eta_x$ and $\eta_y$ denote the derivatives of $\eta$ with
respect to $x$ and $y$. We take $\epsilon$=0.3, $w_1$=$w_2$=$dx$,
$\tau_1$=$dt$, and $\tau_2$=3$dt$ in our simulated results. The
ratio $D_\xi/D_\phi$=0.33 is the relative rate of process (b) with
respect to process (a). The interaction constants $\lambda$ and
$\lambda'$ are 2.34$dx^2$ and 2.225, respectively. All our
simulated results are robust enough against changes of the
parameters.

We use an adaptive mesh refinement method\cite{pfa} to perform
effectively our phase-field simulations in two dimensional space.
The adaptive mesh refinement enables us to simulate a large
spatial scale ($\sim\mu$m) within an acceptable computational
time, and to get enough details about the regions where the phase
fields change drastically without adding too much computational
time.

\begin{figure}[htbp]
\includegraphics[width=8.5cm]{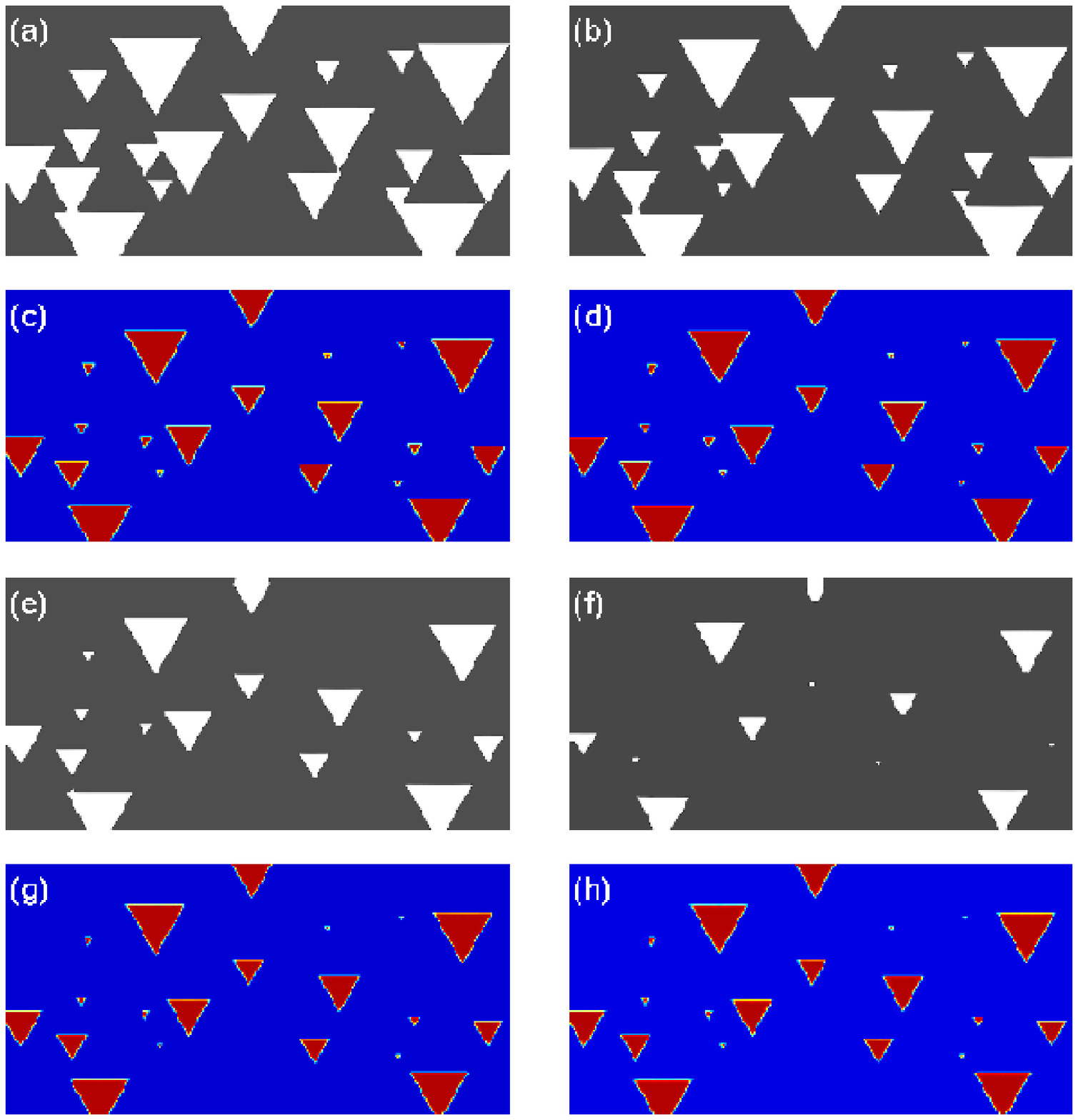}
\caption{(color online). Simulated images of the $\phi$ and $\xi$
fields within the two-phase-field model. White triangles denote
the $\phi$=1 islands and black background the $\phi$=-1 region in
(a), (b), (e), and (f), and red (grey) triangles denote the
$\xi$=1 islands and blue (black) background the $\xi$=-1 region in
(c), (d), (g),  and (h). The time interval is 200, 400, or 300
time steps between (a) and (b), (b) and (e), or (e) and (f),
respectively, where 100 time steps are equal to 8
seconds\cite{l3}. } \label{fig2}
\end{figure}
\begin{figure}[h]
\includegraphics[width=8.6cm]{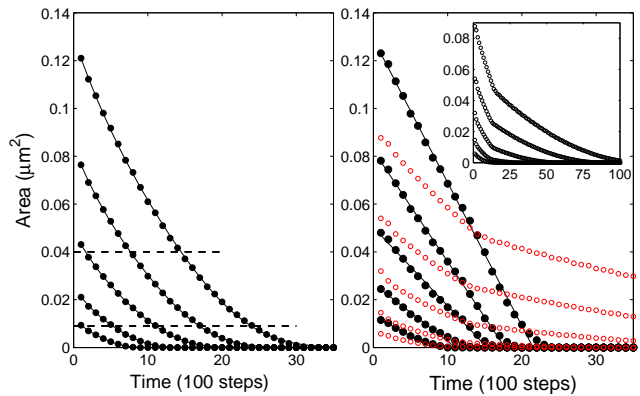}
\caption{(color online). Simulated decay behaviors of five
$\phi$=1 islands with initial areas $S_0$=0.1210, 0.0764, 0.0431,
0.0211, and 0.0093 $\mu m^2$ (from top to bottom) within the
single phase-field model (left) and the two-phase-field model
(right). In the left panel, the circle indicates the simulated
island areas and the curve is fitted with the function
$S_0\exp(-\alpha t- \beta t^{3.5})$. In the right panel, the
filled (hollow) circles indicate the simulated areas of $\phi$=1
($\xi$=1) islands and the lines are linearly fitted $\phi$=1
results. The inset shows the $\xi$=1 areas over the whole
simulation time. } \label{fig3}
\end{figure}

{\it Main findings through simulation} The simulated images are
presented in Fig. \ref{fig2}. The whole time sequence of the
7$\times$7$\rightarrow$``1$\times$1'' phase transition is shown in
the eight panels (a-h), among which (a), (b), (e), and (f) for
$\phi$ and (c), (d), (g), and (h) for $\xi$. At the beginning, the
$\xi$=1 region is always enclosed in a larger $\phi$=1 region, as
shown in the four panels (a)-(d). The $\xi$=1 region must be right
in size with respect to the $\phi$=1 region in order to achieve
the linear area decay behavior of the 7$\times$7 regions. Too
small $\xi$=1 region is not enough to drive the decay from the
exponential behavior, and on the other hand too large $\xi$=1
region overdrives the area decay from the linear behavior. The
right area difference satisfies the condition that
$S^0_{\phi}-S^0_{\xi}$ is a linear function of $S^0_{\phi}$. This
can be understood by considering the fact that when the sample is
quenched, the slow faulted stacking lags the formation of the
dimers and corner holes.

The area decays for five islands are presented in the right panel
of Fig. \ref{fig3}. The filled circles are the areas of $\phi$=1
islands and the hollow ones those of $\xi$=1 regions. The area
decays are linear with time and the decay rates are 0.0060,
0.0046, 0.0031, 0.0021 and 0.0014 ($\mu$m$^2$ per 100 time steps)
for the initial areas: 0.1210, 0.0764, 0.0431, 0.0211, and 0.0093
$\mu$m$^2$, respectively. It can be concluded that the area decay
rate of a 7$\times$7 island increases with its initial area, being
independent of its current area. An initially larger island always
decays faster than an initially smaller one even when their sizes
become equivalent to each other. The area decay rate of a
7$\times$7 island remains the same until the island disappears.

If using only one phase-field variable, we get area decay results
shown in the left panel of Fig. \ref{fig3}. The filled circles
show the areas of five islands with the same five initial areas as
those of the two-phase-field model. They can be fitted with the
function $S_0\exp(-\alpha t- \beta t^{3.5})$, with the parameters
$\alpha$ and $\beta$ decreasing with $S_0$. Area decay rates of
different 7$\times$7 islands increase with their current areas,
not directly depending on their initial areas. The rate is 0.0047
$\mu$m$^2$ per 100 time steps at $S$=0.04$\mu$m$^2$ (upper dash
line), and reduces to 0.0024 $\mu$m$^2$ per 100 time steps at
$S$=0.01$\mu$m$^2$ (lower dash line). This behavior is not
compatible with the experimental LEEM results\cite{l3}, and
therefore the two phase-field variables are necessary to
explaining the experiment.

\begin{figure}[htbp]
\includegraphics[width=8.5cm]{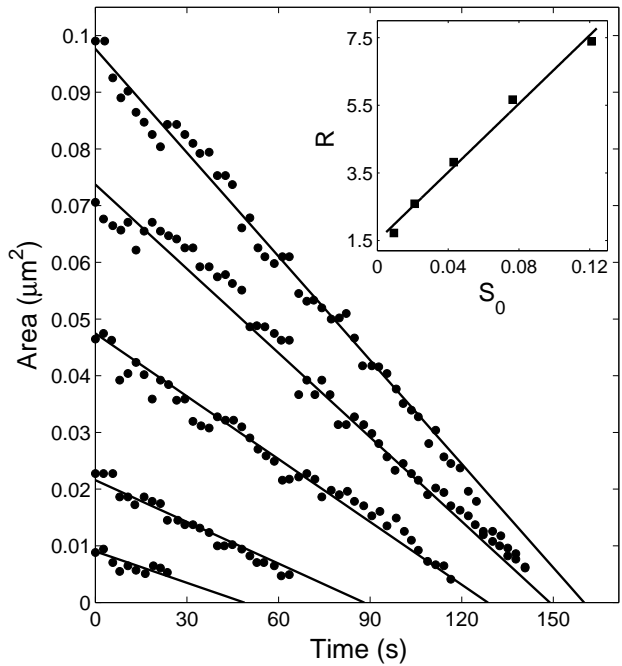}
\caption{The two-phase-field simulated area decay (lines) of five
7$\times$7 islands compared with experimental LEEM data
(dots)\cite{l3}. Inset: the simulated decay rate ($R$ in units of
$10^{-4}\mu m^2/s$) vs the initial area $S_0$ ($\mu m^2$), where
the squares, derived from Fig. 3, are fitted with a linear
function. } \label{fig4}
\end{figure}
{\it Comparison with experiment} In Fig. \ref{fig4} we compare our
simulated area decay results ($\phi$=1) in terms of the
two-phase-field model with experimental data concerned. In the
inset we present the relation of the area decay rate ($R$) with
the initial area ($S_0$). We obtain a linear function,
$R=0.00015+0.0051 S_0$. With this function, the area decay rates
are obtained by comparing the simulated island areas with
experimental ones, or vice versa. The experimental island area $S$
can be described with a linear function of the time $t$,
$S=S_0-Rt$. Because initial island area $S_0$ can be determined
more accurately in LEEM experiment, we obtain $S_0$ by fitting
experimental area data. The fitted $S_0$ are 0.0976, 0.0737,
0.0475, 0.0216, and 0.0090$\mu m^2$, and the corresponding area
decay rates are 6.4, 5.2, 3.9, 2.6, and 2.0$\times 10^{-4}\mu
m^2/s$. The theoretical linear behaviors are in good agreement
with the experimental data\cite{l3}. Therefore, our theoretical
model is very good for understanding the experimental results.

{\it Conclusion} We have proposed a natural two-speed model for
the phase dynamics of the phase transition of Si(111) 7$\times$7
islands to unreconstructed high temperature ``1$\times$1'' phase.
We formulate the phase dynamics by using phase-field method and
adaptive mesh refinement technique. Our simulated results show
that a 7$\times$7 island decays through step flow and the
triangular shape is kept all the time. The decay rate of the
island area is shown to be approximately a constant increasing
with the initial area only. The LEEM experiments are explained
quantitatively. This in return supports our conclusion: the corner
holes and dimer chains are broken first and the remedying of the
stacking-fault is slower and takes longer time. Therefore, the
phase dynamics is elucidated. This phase-field theory is a
reliable approach to studying the phase dynamics of general
surface phase transitions.

\begin{acknowledgments}
This work is supported  by Nature Science Foundation of China
(Grant Nos. 10774180, 90406010, and 60621091), by Chinese
Department of Science and Technology (Grant No. 2005CB623602), and
by the Chinese Academy of Sciences (Grant No. KJCX2.YW.W09-5).
\end{acknowledgments}

\end{document}